\documentclass[aps,groupedaddress,nofootinbib]{revtex4}

\usepackage[bookmarks=true,bookmarksopen=true,pdfhighlight=/I,pdfpagemode=UseOutlines]{hyperref}
\usepackage[latin1]{inputenc}
\usepackage{graphicx}
\usepackage{amssymb,amsmath}
\usepackage[usenames,dvipsnames]{color}
\usepackage{hyperref}
\newcommand{\be}{\begin{equation}}
\newcommand{\ee}{\end{equation}}
\newcommand{\bea}{\begin{eqnarray}}
\newcommand{\eea}{\end{eqnarray}}	
\newcommand{\nn}{\nonumber\\}
\newcommand{\ba}{\begin{array}}
\newcommand{\ea}{\end{array}}

\newcommand{\rfv}{|0\rangle_f}
\newcommand{\lfv}{_f\langle 0 |}
\newcommand{\rmv}{| 0 \rangle}
\newcommand{\lmv}{\langle 0 |}
\newcommand{\vx}{\vec{x}}

\newcommand{\vk}{\vec{k}}

\newcommand{\sst}{\sin^2 \theta}

\newcommand{\w}{\omega}

\begin{document}

\title[Flavour Vacua \& SUSY]{Quantum Gravity, Flavour Vacua and Supersymmetry}

\author{Nick E. Mavromatos}
\affiliation{CERN, Theory Division, CH-1211  Geneva 23, Switzerland; \\On leave from: King's College London, Department of Physics, Strand, London WC2R 2LS, UK.}

\begin{abstract}
 I review a novel and non-perturbative way of breaking Supersymmetry in the flavour sector of electrically neutral particles, as a result of flavour mixing in the presence of stringy quantum gravity space-time foam backgrounds that violate Lorentz and CPT symmetries. In these models, part of the mixing may itself be induced dynamically by the interactions of the flavoured particles with the space-time foam.

\end{abstract}
\maketitle                   

\section{Introduction}

The most promising framework for a quantum theory of gravity is string theory,
particularly in its non-perturbative formulation known as M-theory.
This contains solitonic configurations such as D-branes~\cite{polchinski}, including
D-particle defects in space-time. One of the most challenging problems
in quantum gravity is the description of the vacuum and its properties.
At the classical level, the vacuum may be analyzed using the tools of
critical string theory. However, we have argued~\cite{emn1} that a consistent
approach to quantum fluctuations in the vacuum, the so-called `space-time
foam', needs the tools of non-critical string theory. As an example,
we have outlined an approach to this problem in which D-branes
and D-particles play an essential role~\cite{Dfoam}.
According to this class of models,  our Universe, perhaps after appropriate compactification, is represented  as a Dirichlet three-brane (D3-brane), propagating in a bulk
space-time punctured by D-particle defects. Since an isolated D-particle cannot exist~\cite{polchinski},
because of gauge flux conservation, the presence of a D-brane is essential.
As the D3-brane world moves
through the bulk, the D-particles cross it. To an observer on the D3-brane the model looks
like `space-time foam' with defects `flashing' on and off: this is the
structure we term `D-foam'. Matter particles are represented in this scenario by open strings
whose ends are attached to the D3-brane. They can interact with the D-particles
through splitting and capture of the strings by the D-particles, and subsequent
re-emission of the open string state. The process is characterised by creation of intermediate string states, stretched 
between the D-particle and the D3-brane. This
set-up for D-foam can be considered either in the context of type-IIA string
theory~\cite{emnnewuncert}, in which the D-particles are represented by
point-like D0-branes, or in the context of the phenomenologically more
realistic type-IIB strings~\cite{li}, in which case the D-particles
are modelled as D3-branes compactified around spatial three-cycles (in the simplest scenario),
since the theory admits no D0-branes.
These models have a rich phenomenology, making concrete predictions  in both astroparticle~\cite{mavro_review2009} and terrestrial particle physics~\cite{omega}, in view of their Lorentz and CPT Violating properties.

In \cite{mavrosarkar} we have embarked in an analysis of other effects of D-foam, specifically its effects on flavoured particles. In a multiflavour field theory with mixing, it has been suggested by Vitiello and Blasone (BV)~\cite{blasone} that the physically ``correct'' vacuum, whose excitations describe the observed flavour states, is not the mass-eigenstate one but rather the unitarily inequivalent, ``flavour vacuum'', which is orthogonal to the former  in the thermodynamic limit of infinite volume of the respective field theory. 
The physically observed flavoured particles, \emph{e.g.} $\nu_{e,\mu,\tau}$ neutrinos, are states in the  Fock space of the Flavour vacuum, which, in the thermodynamic limit, are orthogonal to the mass-eigenstate Fock-space states. 
In our D-foam model, for reasons of charge conservation~\cite{Dfoam,emnnewuncert,li}  it is the electrically neutral particles that interact 
predominantly with the foam in the sense of being captured and suffering splitting 
when interacting with a D-particle defect. In \cite{mavrosarkar,mstsusy} we discussed how the  
 mixing itself may be induced by the interaction of the mater excitations with the quantum fluctuating D-particles, and we shall review this issue briefly here. An important feature of D-foam is its Lorentz violating nature, which arises because of the recoil of the D-particles during their non-trivial interactions with the open string matter states.  Mixing, in the presence of Lorentz violation implies that the quantisation of the resulting field theory can be performed by a Fock space of the flavour rather than the mass eigenstates, whose ground state is the so called ``flavour vacuum''~\cite{blasone}. The latter is unitarily inequivalent 
 to the mass-eigenstate one. In our D-foam model, the interactions of neutral flavoured fermions with the D-foam lead to the formation of Lorentz and CPT Violating condensates of the fermions in the flavoured vacuum, which break target space Supersymmetry. The breaking is non perturbative, and leads to different equations of state between bosonic and fermionic degrees of freedom. The latter behave as dust-like fluids, while the former resemble a cosmological constant -like (dark energy) fluid. 

\section{Supersymmetry Breaking via Flavour-Vacua: The flavoured Wess-Zumino Model}

In \cite{mstsusy}, whose results we shall review here,  we demonstrated these features in the simplest, but quite informative,  supersymmetric model, a Wess-Zumino multi-flavoured model with mixing, which we view as a low-energy effective field theory of our brane microscopic models. A similar analysis, but not from the point of view of D-foam, nor discussing the equations of state of the constituent fluids, has also appeared in \cite{capolupo}, independently and contemporaneously with our work.  
The Lagrangian of the model reads:
\begin{equation}\label{lag}
\mathcal{L}_{\rm WZ} = \bar{\Psi}_f(i\partial\!\!\!/-M_F)\Psi_f  +  \partial_{\mu}\mathcal S_f \partial^{\mu}\mathcal S_f -\mathcal S_f M_b\mathcal S_f   + \partial_{\mu}\mathcal P_f \partial^{\mu}\mathcal P_f -\mathcal P_f M_b \mathcal P_f 
\end{equation} 
where  $f=A,B,$ is a flavour index, and the mass matrices for fermions  $M_F$ and bosons $M_b$ contain off diagonal terms, representing mixing.  The fields $\Psi_f$, $S_f$ and $P_f$ (with $f=A,B$) are respectively a Majorana spinor,
a scalar and a pseudo-scalar of definite flavour $A$ or $B$. In our case of D-foam, as we shall discuss below, the mass matrices $M_{F,b}$  of the above WZ model, with non-trivial off-diagonal  elements in flavour space, are viewed as the result of the formation of condensates from microscopic four particle interactions. 
Following the BV formalism \cite{blasone}, we introduce the operator $G_{\theta}(t)$, connecting fields between the flavour and mass representations 
$G_{\theta}(t)=e^{\theta \int d\vx \left(X_{12}(x)-X_{21}(x)\right)}$
with $X_{12}(x)\equiv \frac{1}{2}\psi^\dagger_1(x)\psi_2(x)+i \dot{S}_2(x) S_1(x)+i \dot{P}_2(x) P_1(x)$. The \textit{flavour vacuum} is then defined as  $\rfv\equiv G^{\dagger}_{\theta}(t)\rmv $ 
where $\rmv$ is the usual vacuum defined by $a^r_i(\vk)\rmv=b_i(\vk)\rmv=c_i(\vk)\rmv=0$ ($i=1,2$).
In order to study the properties of the \textit{flavour vacuum}, and argue on the induced supersymmetry breaking, wel calculate the \textit{flavour} vacuum expectation value
of the stress-energy tensor, namely
$\lfv T_{\mu\nu}(x) \rfv $. Appropriate subtraction of the zero-mixing contributions is understood, in order to obtain a vanishing result in that case, consistent with the Lorentz invariance in the absence of mixing. 
After straightforward but tedious calculations, one obtains  a vanishing result for the off diagonal space-time components of the stress tensor, 
in the momentarily co-moving rest frame of the fluid, and the following non-trivial results for the energy density $\rho$ and pressure 
$p$ of the total fluid, containing both bosonic and fermionic contributions~\cite{mstsusy}: 
\begin{eqnarray}\label{e1}
\rho &=& \lfv T_{00}(x)\rfv=\sst \frac{(m_1-m_2)^2}{2 \pi^2}\int_0^\Lambda dk\;k^2\left(\frac{1}{\w_1(k)}+\frac{1}{\w_2(k)}\right) 
 =\sst \frac{(m_1-m_2)^2}{2 \pi^2}f(\Lambda) \nn
 p&=&\lfv T_{ii}(x)\rfv=-\sst \frac{m_1^2-m_2^2}{2 \pi^2}\int_0^\Lambda dk\; k^2 \left(\frac{1}{\w_1(k)}-\frac{1}{\w_2(k)}\right)
 =\sst \frac{m_1^2-m_2^2}{2 \pi^2}g(\Lambda) 
\end{eqnarray}
where no sum over $i$ is implied and $\omega_j = \sqrt{k^2 + m_j^2}$,  $j=1,2$. 
The integrals in (\ref{e1}) have been evaluated analytically in ~\cite{mstsusy}. The resulting functions $f(\Lambda)$ and $g(\Lambda)$ depend on the UV cutoff $\Lambda$ that regularizes infinities 
in the flat space-time theory. The function $f(\Lambda) > 0$ and, thus, the energy is \emph{positive}. This is not surprising, since we know that in a supersymmetric field-theory model the state with the lowest energy is unique, and corresponds to the usual vacuum $\rmv$, for which $\lmv T_{00} \rmv =0$,
all other states having positive energies. The difference in energy of the ``flavour vacuum'' from the mass-eigenstate one is due to the
condensation of massive flavoured particle-antiparticle states~\cite{blasone,capocosmo,mavrosarkar}.
The choice of this vacuum leads therefore to a \textit{breaking of supersymmetry}, in a novel way, which is explored further by considering the equations of state $w_{B,F}=p_{B,F}/\rho_{B,F}$ of the individual bosonic (B) and fermionic (F) fluid constituents~\cite{mstsusy}:
$
w_B=\frac{\lfv: \sum_{i=1,2}T_{ii}^{S_i}(x):\rfv}{\lfv: \sum_{i=1,2}T_{00}^{S_i}(x):\rfv}
					=\frac{\lfv: \sum_{i=1,2}T_{ii}^{P_i}(x):\rfv}{\lfv: \sum_{i=1,2}T_{00}^{P_i}(x):\rfv}=-1
$
and
$
w_F=\frac{\lfv: \sum_{i=1,2}T_{ii}^{\psi_i}(x):\rfv}{\lfv: \sum_{i=1,2}T_{00}^{\psi_i}(x):\rfv}=0~,$ where $: \dots :$ denotes appropriarte normali ordering. This result means that the bosonic contribution to the vacuum condensate
by itself would lead to a \emph{pure cosmological constant-type} equation of state~\cite{capolupo,mavrosarkar}.
This behaviour of the fluid is mitigated by the contribution of the Majorana fermionic fluid, that has a \emph{vanishing} pressure.
One is tempted to  interpret the zero total pressure of the fermionic fluid as a result of an extra contribution, as compared with the bosonic case, due to the Pauli exclusion principle, the so called degeneracy pressure. 
The latter is a positive contribution to the total pressure of a fermion fluid which is due to the extra force one has to exert in order to 
form a condensate of fermions in a small (local) region of space time. When two fermions (of the same flavour) are squeezed too close together, the exclusion principle requires them to have opposite spins in order to occupy the same energy level. To add another fermion of this flavour to a given volume (as required by the formation of a condensate) requires raising the fermion's energy level, and this requirement for energy to compress the material appears as a (positive) pressure.

\section{Embedding to Microscopic D-foam Models: Flavour-vacua Condensates and Mixing} 

The above considerations pertain to free Wess-Zumino models with mixing. One may embed such models into our microscopic D-foam framework by considering the mass matrix of (\ref{lag}) as being induced by a \emph{vacuum condensate} due to some contact four particle interactions. Our supersymmetric string theory D-foam models contain (upon compactification and projection on the four space-time dimensional brane worlds) in their low energy limits a plethora of such four-particle interactions. For instance, one finds the following four-multiflavour-fermion 
interaction terms in the low-energy field theory~\cite{li,antoniad}: 
$ \int d^4x \,{\rm det}(e)\, \frac{\eta}{M_s^2} V^c \sum_{i,j} \sum_{a,b=L, R} \mathcal{G}_{a b} \overline{\psi_i}^{a} \gamma^\mu \psi_{i\,a}  \, \overline{\psi_j}^{\prime \,b} \gamma^\mu \psi^\prime_{j\,b}$, 
where ${\rm det}(e) $ is the vierbein determinant,  $V^c$ denotes the compactitication volume element to four space-time dimensions, in units of the string length $M_s^{-1}$; $\mathcal{G}_{ab}$ are numerical coefficients depending on the (square of the) couplings of the particular interactions, and are of order $O(1)$ as far as the string scale $M_s$ is concerned, and $L,R$ denote the appropriate chirality of the spinors $\psi_i, \psi'_j$. The constant $\eta$ depends on the type of D-foam considered`\cite{li,mstsusy}. The combination $\kappa = 8\pi \frac{V^c}{M_s^2} $
defines the gravitational coupling of the four-dimensional effective theory. Four boson interactions also exist, in agreement with target space supersymmetry.  In addition to these contributions, there are also four-fermion terms of axial-current type, induced gravitationally, as a result of the existence of  fermionic torsion contributions to the gravitational spin connection  of the fermion field theory coupled to gravity (Einstein-Cartan theory)~\cite{mercuri}. The corresponding terms are generic and include both charged and neutral fermions, since gravity coupling is universal among particles. These are exact (from a quantum gravity point of view) terms, given that they are induced by the non-propagating torsion field in the path integral of quantum gravity. For  generic spin-$\frac{1}{2}$ fermions,  $\Psi_i$,  such torsion-induced terms assume the form~\cite{mercuri}: 
$ \int d^4 x\, {\rm det}(e) \frac{3}{16}\kappa \sum_{i,j}(\overline{\Psi_i} \gamma^a \gamma_5 \Psi_i )(\overline{\Psi_j} \gamma^a \gamma_5 \Psi_j),$ with $ \gamma_5 = i  \gamma^0 \gamma^1 \gamma^2 \gamma^3$.

Let us concentrate from now on to neutral flavoured fermions, which suffices for the purposes of the talk. In other words, we treat the fermions $\psi $, $\psi^\prime$ and $\Psi$ as referring to neutrino species, which denote generically by $\psi_i$, $i=1, \dots N$ the number of flavours. Combining the four fermion interactions  with the gravitational and fermion kinetic parts of the effective action, and using appropriate Fierz identities, we write for the effective low-energy action:
\begin{eqnarray}\label{final4fermi}
&& S_{\rm G} = \frac{1}{\kappa} \int d^4 x \,{\rm det}(e) \left(\Lambda -\frac{1}{2} e^\mu_{\,a }e^\nu_{\,b} \widehat{R}_{\mu\nu}^{\,\,\,a b}(e) + \sum_{j}\frac{i \kappa}{2} e^\mu_{\, a} [\overline{\psi_j} \gamma^a \widehat{\mathcal{D}}_\mu\, \psi_j - \overline{\widehat{\mathcal{D}}_\mu \psi_j} \gamma^a \psi_j ] \right. +  \nonumber \\ 
 && 
 \left. \kappa^2 (\frac{3}{16} - \eta \mathcal{G} ) [ (\overline{\psi_i} \psi_j )\, (\overline{\psi_j} \psi_i) 
  - (\overline{\psi_i} \gamma_5 \psi_j )\, (\overline{\psi_j} \gamma_5 \psi_i)] \, + \right. \nonumber \\ 
&&  \left.  + \, \frac{1}{2}\,\kappa^2 (\frac{3}{16} + \eta \mathcal{G} )  \left[(\overline{\psi_i} \gamma^\mu \psi_{j} ) \, (\overline{\psi_j } \gamma^\mu \psi_{i})  + (\overline{\psi_i} \gamma_5 \gamma^\mu \psi_{j} ) \, (\overline{\psi_j } \gamma_5 \gamma^\mu \psi_{i})\right] \right)~, 
 \end{eqnarray}
where,   for brevity and concreteness, we have assumed  the existence of a single interaction with (square) coupling $\mathcal{G} > 0 $ due to  intermediate string states stretched between D-particle and brane world. 
The hatted notation denotes terms defined without any fermionic torsion, \emph{i.e.} standard General relativity covariant derivatives and curvature tensors. We use the first order formalism of gravity, involving vierbein $e^a_\mu$ and  
torsion-free connection $\omega^{ab}_\mu(e)$, as appropriate for the coupling of fermionic matter to gravity.
The cosmological constant $\Lambda$  in (\ref{final4fermi})  accounts for vacuum energy contributions from bulk D-particles onto the brane world in the models of D-foam~\cite{Dfoam}. These contributions are of mixed sign, due to both attractive and repulsive forces, thereby leading to the possibility of instabilities of the mass eigenstate vacuum in epochs where the bulk concentration of the D-particle is such that the brane vacuum energy is negative. The formation of condensates in the flavour vacuum may restore stability~\cite{mstsusy}, given that such condensates induce positive contributions to the vacuum energy that may cancel or overcome the negative ones. 
Below we give plausibility arguments as to how this can be achieved~\cite{mstsusy}. 
The Lorentz-invariance violation of the D-foam background allows the formation of vector condensates in (\ref{final4fermi}), which would otherwise be forbidden. Due to the assumed isotropy of the D-foam~\cite{Dfoam,mavrosarkar},  it is only the temporal component of the condensate that can acquire a flavour-vacuum expectation value. In the case of chiral fermions like the neutrinos of the standard model, such 
components  correspond to fermion number, which is known to be non zero in the flavour vacuum formalism~\cite{blasone,mavrosarkar}.
Indeed, consider the vector current $J^\mu = \sum_{i} \overline{\psi}_i \gamma^\mu \psi_i $. The temporal component $J^0$ is just the fermion number operator. The latter is known to be non-trivial in flavour-vacuum formalism, in both flat~\cite{blasone} and de Sitter space times~\cite{mavrosarkar}, thereby leading to effective Lorentz- (and CPT-)violating interactions of neutrinos~\cite{kostel}, as a result of the propagation of the latter in their own condensates.  This is what we call the \emph{gravitational MSW effect}. These condensates contribute \emph{positively} to the vacuum energy of (\ref{final4fermi}), and thus may cancel potentially unstable $\Lambda < 0$ contributions. 

If right handed fermions are allowed in the theory (\ref{final4fermi}), the formation of scalar condensates, \emph{e.g}. due to attractive forces in the case $\eta \mathcal{G} > 3/16$, lead to \emph{induced mixing} 
(from the off-diagonal elements of the condensate in flavour space, $i \ne j$) of a mass matrix type, while the diagonal elements induce mass contributions. In these sense, the mass matrices of the WZ model (\ref{lag}) may be seen as the result of such dynamically generated condensates.  The formation of such condensates in a de Sitter space time does not require strong coupling, unlike the situation in Minkowski space. Indeed, such four fermion attractive terms have been discussed in connection with chiral symmetry breaking in the early universe, and induced dark energy in the late eras, 
but from a different perspective than ours~\cite{neubert}.  In our approach, as we have discussed, neutrino condensates in the flavour vacuum do also contribute to the late-times Universe acceleration, in the form of a dark energy fluid~\cite{capocosmo,mavrosarkar}, but the equations of state of fermions and bosons are different, which is an additional way for breaking supersymmetry in initially supersymmetric models, as those obtained from realistic string theories.

\section*{Acknowledgements}

I wish to thank G. Zoupanos for the invitation to the Corfu School 2010.
This work is partially supported by the European Union
through the Marie Curie RTN network \emph{UniverseNet}
(MRTN-2006-035863).


\begin{thebibliography}{[1]}


\bibitem{polchinski} J. Polchinski, \emph{String Theory}, Vol. 2 (Cambridge University Press, 1998).


\bibitem{emn1} J.~R.~Ellis, N.~E.~Mavromatos and D.~V.~Nanopoulos,
Phys.\ Lett.\ B \textbf{293}, 37 (1992);
\textit{A microscopic Liouville arrow of time}, Invited review for the
special Issue of \textit{J.\ Chaos Solitons Fractals}, Vol.\ 10, p.~345-363
(eds. C. Castro amd M.S. El Naschie, Elsevier Science, Pergamon 1999). 


\bibitem{Dfoam} J.~R.~Ellis, N.~E.~Mavromatos and M.~Westmuckett,
Phys.\ Rev.\ D \textbf{70}, 044036 (2004);
\emph{ibid.} {\bf 71}, 106006 (2005)~.


\bibitem{emnnewuncert} J.~R.~Ellis, N.~E.~Mavromatos and D.~V.~Nanopoulos,
  Phys.\ Lett.\  B {\bf 665}, 412 (2008).


  \bibitem{li} T.~Li, N.~E.~Mavromatos, D.~V.~Nanopoulos and D.~Xie,
  Phys.\ Lett.\  B {\bf 679}, 407 (2009).



\bibitem{mavro_review2009} N.~E.~Mavromatos,
  Int.\ J.\ Mod.\ Phys.\  A {\bf 25}, 5409 (2010)  and references therein.


\bibitem{omega} J.~Bernabeu, N.~E.~Mavromatos and J.~Papavassiliou,
  Phys.\ Rev.\ Lett.\  {\bf 92}, 131601 (2004);
N.~E.~Mavromatos,
  Found.\ Phys.\  {\bf 40}, 917 (2010) and references therein.


      

\bibitem{mavrosarkar} N.~E.~Mavromatos and S.~Sarkar,
  New J.\ Phys.\  {\bf 10}, 073009 (2008);
N.~E.~Mavromatos, S.~Sarkar and W.~Tarantino,
  Phys.\ Rev.\  D {\bf 80}, 084046  (2009) .
  
 \bibitem{blasone}
  M.~Blasone and G.~Vitiello,
  Annals Phys.\  {\bf 244}, 283 (1995)
  [Erratum-ibid.\  {\bf 249}, 363  (1996) 363].
  
\bibitem{mstsusy} N.~E.~Mavromatos, S.~Sarkar and W.~Tarantino,
  arXiv:1010.0345 [hep-th].
  
 \bibitem{capolupo} A.~Capolupo, M.~Di Mauro and A.~Iorio,
  arXiv:1009.5041 [hep-th].
  

\bibitem{capocosmo} A.~Capolupo, S.~Capozziello and G.~Vitiello,
  Phys.\ Lett.\  A {\bf 363}, 53 (2007);
  \emph{ibid.} {\bf 373}, 601 (2009).

    
  \bibitem{antoniad} 
I.~Antoniadis, K.~Benakli and A.~Laugier,
  JHEP {\bf 0105}, 044 (2001);
D.~Lust \emph{et al}.
  Nucl.\ Phys.\  B {\bf 828}, 139 (2010).
  
    
  \bibitem{mercuri}  S.~Mercuri,
  Phys.\ Rev.\  D {\bf 73}, 084016 (2006)  and references therein.
  
\bibitem{kostel}  V.~A.~Kostelecky and M.~Mewes,
  Phys.\ Rev.\  D {\bf 70}, 076002 (2004); S.H.S.~Alexander,
  arXiv:0911.5156 [hep-ph].
  

  \bibitem{neubert} F.~Giacosa, R.~Hofmann and M.~Neubert,
  JHEP {\bf 0802}, 077 (2008);
  S.~Alexander, T.~Biswas and G.~Calcagni,
  Phys.\ Rev.\  D {\bf 81}, 043511 (2010)
  [Erratum-ibid.\  D {\bf 81}, 069902 (2010)].
  
\end{thebibliography}
\end{document}